# Frequency-Domain Chromatic Dispersion Equalization Using Overlap-Add Methods in Coherent Optical System

Tianhua Xu[1,2,3], Gunnar Jacobsen[2], Sergei Popov[1], Marco Forzati[2], Jonas Mårtensson[2], Marco Mussolin[4], Jie Li[2], Ke Wang[1], Yimo Zhang[3], Ari T. Friberg[1]

**Summary**

The frequency domain equalizers (FDEs) employing two types of overlap-add zero-padding (OLA-ZP) methods are applied to compensate the chromatic dispersion in a 112-Gbit/s non-return-to-zero polarization division multiplexed quadrature phase shift keying (NRZ-PDM-QPSK) coherent optical transmission system. Simulation results demonstrate that the OLA-ZP methods can achieve the same acceptable performance as the overlap-save method. The required minimum overlap (or zero-padding) in the FDE is derived, and the optimum fast Fourier transform length to minimize the computational complexity is also analyzed.

**Keywords.** Frequency domain equalizers (FDEs), overlap-save (OLS), overlap-add (OLA), chromatic dispersion (CD), quadrature phase shift keying (QPSK).

## 1  Introduction

Fiber impairment such as the chromatic dispersion (CD) severely impacts the performance of the high speed optical fiber transmission systems [1,2]. Current systems usually use dispersion compensation fibers (DCFs) to suppress the CD distortion in the optical domain, which increases the cost and deteriorates the nonlinear tolerance of the transmission systems. Coherent detection allows dispersion equalization in the electrical domain, and has become a promising alternative approach to optical dispersion compensation (ODC) [3,4].

Several digital filters have been applied to compensate the CD in the time and the frequency domain [4-6]. Compared to the time-domain fiber dispersion finite impulse response (FD-FIR) and adaptive least mean square (LMS) filters, the frequency domain equalizers (FDEs) have become the more attractive digital filters for channel equalization in the coherent transmission systems due to the low computational complexity for large dispersion and the wide applicability for different fiber distance [4-8]. The fast Fourier transform (FFT) convolution algorithms involving the overlap-save (OLS) and the overlap-add zero-padding (OLA-ZP) methods are traditionally used for the frequency domain equalization in the wireless communication systems [9-12]. Recently, the OLS-FDE employed in the coherent optical communication system was reported, where the received data sequence is divided into small blocks with a certain overlap before they are equalized [13,14].

In this paper, two types of FDEs employing the OLA-ZP FFT convolution methods are investigated to compensate the CD in a 112-Gbit/s non-return-to-zero polarization division multiplexed quadrature phase shift keying (NRZ-PDM-QPSK) coherent optical transmission system [9-12]. In the OLA-ZP equalization, the received data sequence is divided into small blocks without any overlap, but appended with zero-padding. The CD compensation results using the two OLA-ZP methods are compared with the OLS method by evaluating the behavior of the bit-error-rate (BER) versus the optical signal-to-noise ratio (OSNR) as well as the FFT-sizes and the overlap sizes. The minimum value of the overlap (or zero-padding), which is the pivotal parameter in FDE, is evaluated according to the equalized dispersion. Moreover, the optimum FFT-size in FDE is also analyzed to minimize the computational complexity.

**Address of authors:**

[1]Royal Institute of Technology, Stockholm, SE-16440, Sweden

[2]Acreo Swedish ICT AB, Stockholm, SE-16440, Sweden

[3]Tianjin University, Tianjin, 300072, China

[4]University of Padova, Padova, IT-35100, Italy

Email: tianhua@kth.se, tianhuaxu@outlook.com

## 2 Principle of OLS and OLA-ZP methods

### 2.1 Overlap-save method

The schematic of the FDE with overlap-save method is illustrated in Fig. 1 [9,10,13,14]. The received signals are divided into several blocks with a certain overlap, where the block length is called the FFT-size. The sequence in each block is transformed into the frequency domain data by the FFT operation, and afterwards multiplied by the transfer function of the FDE. Next, the data sequences are transformed into the time domain signals by the inverse FFT (IFFT) operation. Finally, the processed data blocks are combined together, and the bilateral overlap samples are symmetrically discarded.

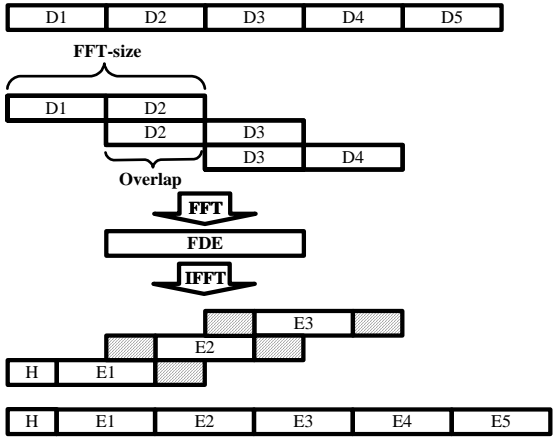

Fig. 1: FDE with OLS method. The parts with slants are to be discarded

The transfer function of FDE is expressed as follows [1],

$$G_c(z,\omega) = \exp(-jD\lambda^2\omega^2 z/4\pi c) \quad (1)$$

where $D$ is the CD coefficient, $\lambda$ is the operation wavelength of the laser, $c$ is the light speed in vacuum, $\omega$ is the angular frequency, and $z$ is the fiber length.

### 2.2 Overlap-add one-side zero-padding method

The structure of the FDE with overlap-add one-side zero-padding (OLA-OSZP) method is shown in Fig. 2 [9-12]. The received data are divided into small blocks without any overlap, and then the data in each block are appended with zeros at one side. To be consistent with the OLS method, the total length of data block and zero padding is called the FFT-size, while the length of zero padding is called the overlap. The zero-padded sequence is transformed by the FFT operation, and multiplied by the transfer function of the FDE. Afterwards, the data are transformed by the IFFT operation. Finally the processed data sequences are combined by overlapping and adding. Note that half of the data stream in the first block is discarded.

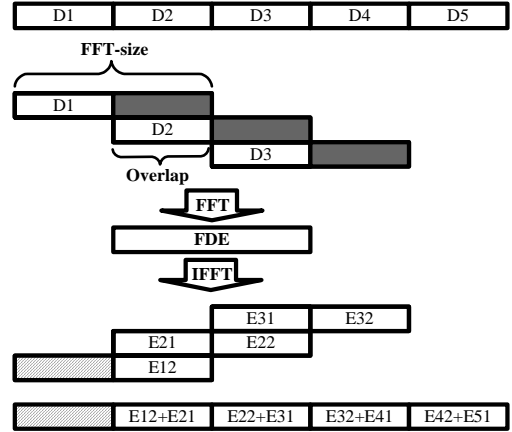

Fig. 2: FDE with OLA-OSZP method; the gray parts mean the appended zeros, and the parts with slants are to be discarded

### 2.3 Overlap-add both-side zero-padding method

The schematic of the FDE with overlap-add both-side zero-padding (OLA-BSZP) method is illustrated in Fig. 3 [9-12]. The received data are also divided into several blocks without any overlap, and then the data in each block are appended with equivalent zeros at both sides. The total length of data block and zero padding is called the FFT-size, and the length of the whole zero padding is called the overlap. The zero-padded sequence is transformed by the FFT operation, and multiplied by the transfer function of the FDE, and then transformed by the IFFT operation. The processed data blocks are also combined together by overlapping and adding. Note that half of the data stream in the first block is discarded.

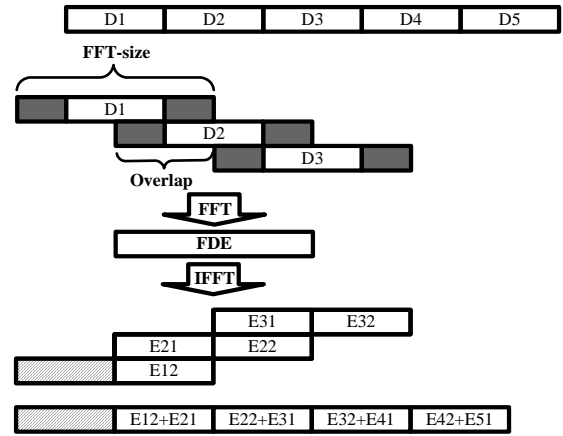

Fig. 3: FDE with OLA-BSZP method; the gray parts mean the appended zeros, and the parts with slants are to be discarded

### 2.4 Minimum overlap in FDE

Actually, the value of the overlap in the OLS method or the zero-padding in the OLA-ZP methods is the pivotal parameter in the FDE determined by the dispersion to be equalized. The FFT-size can be configurable provided it is larger than the overlap (or zero-padding). The required minimum overlap (or ZP) in the OLS or the OLA-ZP method can be calculated from the pulse width broadening (PWB) [1,15],

$$N_P = 2 \times \left\lceil \frac{T_P}{2T} \right\rceil + 2 \qquad (2)$$

$$= 2 \times \left\lceil \frac{1}{\pi c T^2} \sqrt{\pi^2 c^2 T^4 + 4\lambda^4 D^2 z^2} \right\rceil + 2$$

$$T_P = \frac{2}{\pi c T} \sqrt{\pi^2 c^2 T^4 + 4\lambda^4 D^2 z^2} \qquad (3)$$

where $T_P$ is the duration width of a broadened Gaussian pulse, $T$ is the sampling period, and $\lceil x \rceil$ denotes the nearest integer larger than $x$.

Table 1: The minimum overlap in FDE; $N_P$: minimum overlap determined by PWB, $N_S$: minimum overlap determined by CD equalization simulation

| Fiber length (km) | $N_P$ | $N_S$ | $(N_S-N_P)/N_P$ (%) |
|---|---|---|---|
| 20 | 8 | 8 | 0 |
| 40 | 14 | 16 | 14.29 |
| 600 | 158 | 176 | 11.39 |
| 1000 | 260 | 288 | 10.77 |
| 2000 | 518 | 576 | 11.2 |
| 4000 | 1032 | 1152 | 11.63 |
| 6000 | 1546 | 1674 | 8.28 |

The minimum overlap (or ZP) in the FDE for different fiber length is illustrated in Table 1. The CD coefficient of the fiber is 16 ps/nm/km. The value $N_P$ represents the minimum overlap (or ZP) calculated from Eq. (2), and the value $N_S$ represents necessary overlap (or ZP) determined in the dispersion equalization simulation, which will be discussed in the Section 4. We can find that the simulation results achieve a good agreement with the theoretical analysis. It will be demonstrated in our simulation that the performance of CD equalization will degrade drastically, when the overlap (or ZP) in FDE is less than $N_S$. Therefore, the column 4 in Table 1 provides the significantly meaningful information that the theoretical overlap (or ZP) $N_P$ plus about 15% additional supplement can cover the necessary overlap (or ZP) $N_S$ determined in the numerical simulation, which could achieve the satisfactory CD equalization in practical work for the fiber length up to 6000 km.

For a proper overlap (or ZP) value, a large FFT length may be more efficient. However, it will cost more computational complexity and hardware memory resources [16]. The efficient selection of the FFT-size will be discussed later. In our simulation work, the FFT-size is designated as the double of the overlap (or ZP) unless otherwise stated. Therefore, the FDE algorithms can be applied conveniently for equalizing different fiber dispersion only by determining the required FFT-size.

## 2.5 Optimization of FFT-size

From the above analysis, the required overlap (or ZP) depends on the fiber dispersion to be equalized, and any integer, provided larger than the overlap, can be theoretically selected as the FFT-size. However, an optimal FFT-size can be selected to obtain the minimum complexity for frequency domain equalization [16]. The complexity in FDE for different FFT-size using several classical FFT algorithms (such as radix-2 and radix-4 FFT operation) is evaluated by the number of multiplications per symbol (Mul/Sym), which can be calculated as [16,17],

$$N_{Mul} = \frac{N_{FFT} \cdot [6C \cdot \log_2(N_{FFT}) + 3]}{N_{FFT} - N_{Overlap} + 1} \qquad (4)$$

where $N_{FFT}$ is the FFT-size in FDE, $N_{Overlap}$ is the required overlap (or ZP) derived from the fiber dispersion, and $C$ is a positive constant varying for different FFT algorithms. In classical FFT algorithms, $C = 1/2$ corresponds to the radix-2 FFT algorithm (FFT-size equal to power of two), and $C = 3/8$ corresponds to the radix-4 FFT algorithm (FFT-size equal to power of four) [17].

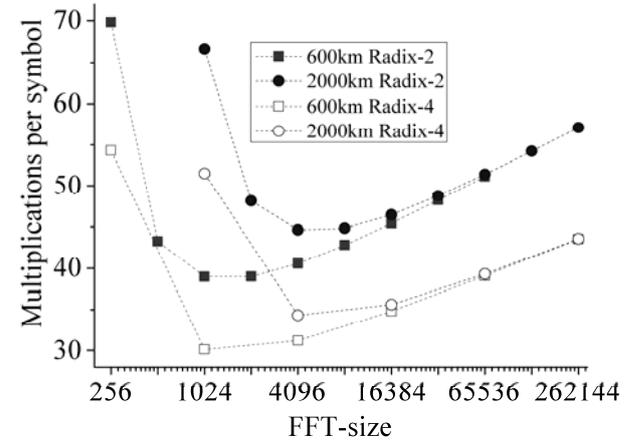

Fig. 4: The complexity for different FFT-size in FDE

The complexity (defined as multiplications per symbol) for 600 km and 2000 km fibers CD equalization using different FFT-size in FDE is shown in Fig. 4. We can see that the optimum FFT-size values to minimize the complexity for 600 km and 2000 km fibers dispersion equalization are 1024 and 4096 respectively in both of the radix-2 and the radix-4 FFT algorithms.

The optimum FFT-size in FDE for different fiber length to minimize the complexity is illustrated in Table 2. We can find that the radix-4 FFT algorithm can achieve a lower complexity than the radix-2 FFT algorithm. Note that the more sophisticated split-radix FFT algorithms can achieve the lowest complexity for FDE with $C \leq 1/3$ in Eq. (4) [16,17].

Table 2: The optimum FFT-size in FDE for different fiber length

| Length (km) | Radix-2 FFT | | Radix-4 FFT | |
|---|---|---|---|---|
| | FFT-size | Mul/Sym | FFT-size | Mul/Sym |
| 20 | 32 | 23.04 | 64 | 18.53 |
| 40 | 64 | 26.35 | 64 | 20.71 |
| 600 | 1024 | 38.98 | 1024 | 30.12 |
| 1000 | 2048 | 41.21 | 4096 | 32.03 |
| 2000 | 4096 | 44.63 | 4096 | 34.33 |
| 4000 | 16384 | 48.02 | 16384 | 36.82 |
| 6000 | 16384 | 49.69 | 16384 | 38.09 |

## 3 NRZ-PDM-QPSK coherent system

The setup of the 112-Gbit/s NRZ-PDM-QPSK coherent transmission system implemented in VPI simulation platform is illustrated in Fig. 5 [18]. The electrical data output from the four 28-Gbit/s pseudo random bit sequence (PRBS) generators are modulated into two orthogonally polarized NRZ-QPSK optical signals by two Mach-Zehnder modulators, and then integrated into one fiber channel by a polarization beam combiner (PBC) to form the 112-Gbit/s PDM-QPSK optical signals. Using a local oscillator (LO) in the coherent receiver, the received optical signals are mixed with the LO laser and transformed into four electrical signals after the photodiodes, which are then digitalized by the analog-to-digital convertors (ADCs) at twice the symbol rate. The CD coefficient in the transmission fiber is 16 ps/nm/km, and the central wavelengths of the transmitter and the LO lasers are both 1553.6 nm. Here the influences of fiber channel attenuation, polarization mode dispersion, phase noise and nonlinear effects are neglected.

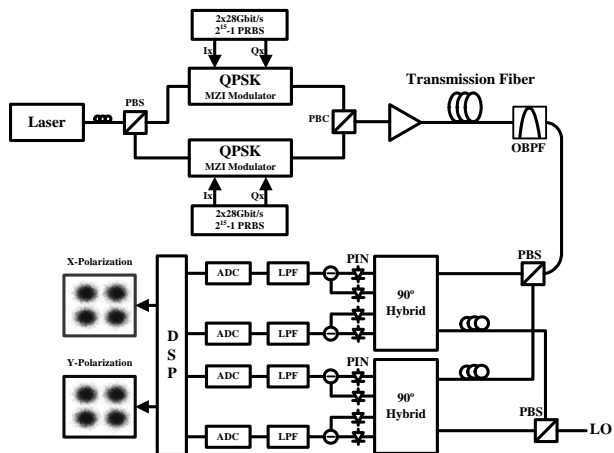

Fig. 5: The 112-Gbit/s NRZ-PDM-QPSK transmission system

## 4 Simulation results

The CD compensation results using the three frequency domain equalization methods are illustrated in Fig. 6.

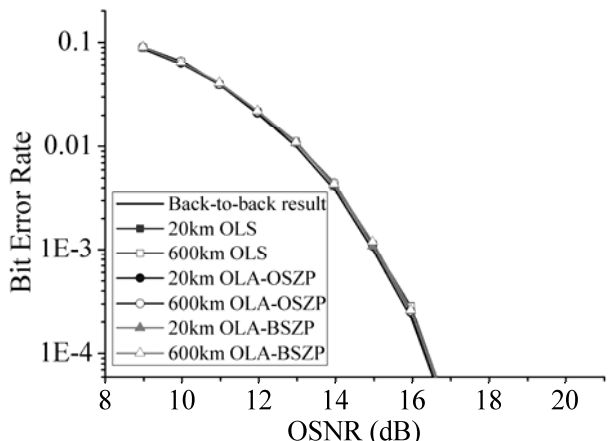

Fig. 6: CD compensation results using OLS and OLA-ZP methods

The results refer to the CD equalization with 16 FFT-size for 20 km fiber and 512 FFT-size for 600 km fiber using OLS, OLA-BSZP and OLA-OSZP methods. The overlap size (or ZP) is all designated as half of the FFT-size. We can see that both of the OLA-ZP methods can provide the same acceptable performance as the OLS method.

Figure 7 and Fig. 8 show the performance of CD compensation for 20 km and 40 km fibers using OLS and OLA-ZP methods with different FFT-sizes and overlaps (or ZP), respectively.

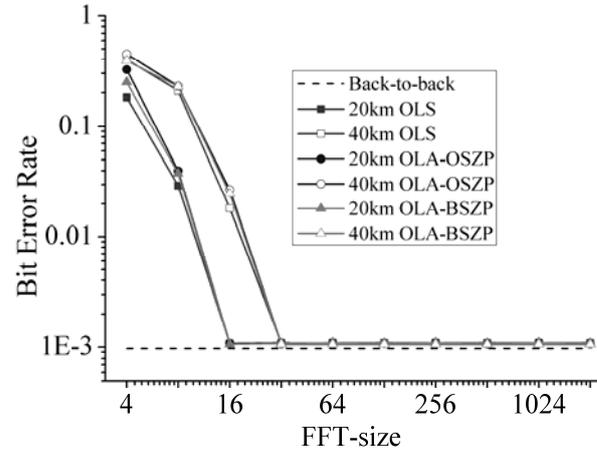

Fig. 7: CD compensation using OLS and OLA-ZP methods with different FFT-sizes at OSNR 14.8 dB; the overlap is half of the FFT-size

From Fig. 7 we can see that for a certain fiber length, the three methods can show stable and converged acceptable performance with the increment of the FFT-size. The critical FFT-size values (16 FFT-size for 20 km fiber and 32 FFT-size for 40 km fiber), actually indicate the required minimum overlap (or ZP) value which are 8 overlap (or ZP) samples for 20 km fiber and 16 overlap (or ZP) samples for 40 km fiber. The similar performance demonstrates that for a fixed overlap (or ZP) value, the maximum compensable dispersion in the OLS method is the same in the OLA-ZP methods.

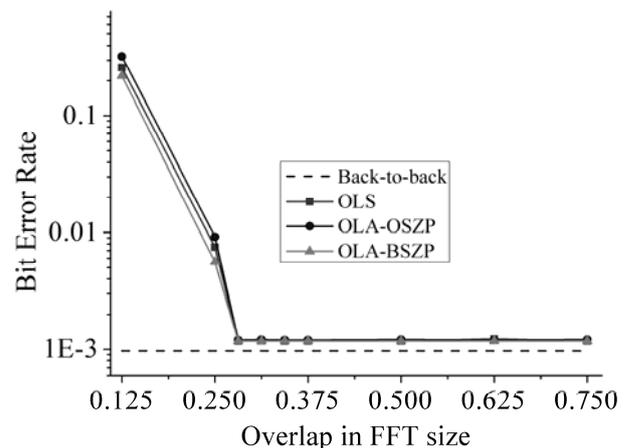

Fig. 8: CD compensation for 4000 km fiber using OLS and OLA-ZP methods with different overlaps at OSNR 14.8 dB; the FFT-size is 4096

We have demonstrated that the overlap (or ZP) is the pivotal parameter in the FDE, and the FFT-size is not necessarily designated as double of the overlap (or ZP). Figure 8 illustrates that with a fixed FFT-size (4096 samples) the three filters are still able to work well for 4000 km fiber, provided the overlap (or ZP) is larger than 1152 samples (1152=4096×9/32), which indicates the required minimum overlap (or ZP) for 4000 km fiber. The minimum overlap (or ZP) in FDE determined from simulation is illustrated in Table 1.

## 5   Conclusions

For the first time to our knowledge, we present the detailed comparative analysis for three types of frequency domain equalization, including overlap-save, overlap-add one-side zero-padding and overlap-add both-side zero-padding methods. They are applied to compensate the CD in the 112-Gbit/s NRZ-PDM-QPSK coherent optical transmission system. Our analysis demonstrates that both the two overlap-add zero-padding methods and the overlap-save method can achieve the same performance in frequency domain CD equalization. The required minimum overlap (or zero-padding) is given out by the analytical expression, and the simulation results show the theoretical overlap plus a 15% additional supplement can provide the acceptable equalization performance. The optimum FFT-size in FDE is also analyzed to obtain the minimum computational complexity. The radix-4 FFT algorithm can be employed to achieve nearly the same lowest complexity as the sophisticated split-radix FFT method.